\documentclass[acmtog]{acmart}

\renewcommand\footnotetextcopyrightpermission[1]{} 
\usepackage{booktabs} 
\usepackage[utf8]{inputenc}
\usepackage{graphicx}
\usepackage{amsmath}
\usepackage{scalerel}
\usepackage{bm}
\usepackage{color,soul}
\graphicspath{ {images/} }

\newcommand{\bp}{\textbf{p}}
\newcommand{\bff}{\textbf{f}}
\newcommand{\bi}{\textbf{i}}
\newcommand{\bo}{\textbf{o}}
\newcommand{\bc}{\textbf{c}}
\newcommand{\bj}{\textbf{j}}
\newcommand{\bz}{\textbf{z}}
\newcommand{\be}{\textbf{e}}
\newcommand{\by}{\textbf{y}}
\newcommand{\bt}{\textbf{t}}

\newcommand{\bx}{\textbf{x}}
\newcommand{\bv}{\textbf{v}}
\newcommand{\br}{\textbf{r}}
\newcommand{\bb}{\textbf{b}}
\newcommand{\bh}{\textbf{h}}

\newcommand{\bW}{\textbf{W}}
\newcommand{\bU}{\textbf{U}}
\newcommand{\bC}{\textbf{C}}
\newcommand{\bF}{\textbf{F}}

\newcommand{\bH}{\textbf{H}}
\newcommand{\bX}{\textbf{X}}
\newcommand{\bY}{\mathrm{\textbf{Y}}}

\setcitestyle{square}

\usepackage[ruled]{algorithm2e} 

\SetAlFnt{\small}
\SetAlCapFnt{\small}
\SetAlCapNameFnt{\small}
\SetAlCapHSkip{0pt}
\IncMargin{-\parindent}

\setcopyright{none}

\begin{document}

\title{Recurrent Transition Networks for Character Locomotion}

\author{Félix G. Harvey}
\orcid{1234-5678-9012-3456}
\affiliation{%
  \institution{Polytechnique Montreal, Mila, Ubisoft Montreal}
  \city{Montreal}
  \state{QC}
  \country{Canada}}
\email{felix.gingras-harvey@polymtl.ca}

\author{Christopher Pal}
\affiliation{%
  \institution{Polytechnique Montreal, Mila, Element AI}
  \city{Montreal}
  \state{QC}
  \country{Canada}}
\email{chris.pal@polymtl.ca}

\renewcommand\shortauthors{Harvey, F. \& Pal, C}

\begin{abstract}
Manually authoring transition animations for a complete locomotion system can be a tedious and time-consuming task, especially for large games that allow complex and constrained locomotion movements, where the number of transitions grows exponentially with the number of states. 
In this paper, we present a novel approach, based on deep recurrent neural networks, to automatically generate such transitions given a \textit{past context} of a few frames and a target character state to reach. We present the Recurrent Transition Network (RTN), based on a modified version of the  Long-Short-Term-Memory (LSTM) network, designed specifically for transition generation and trained without any gait, phase, contact or action labels. We further propose a simple yet principled way to initialize the hidden states of the LSTM layer for a given sequence which improves the performance and generalization to new motions. We both quantitatively and qualitatively evaluate our system and show that making the network terrain-aware by adding a local terrain representation to the input yields better performance for rough-terrain navigation on long transitions. Our system produces realistic and fluid transitions that rival the quality of Motion Capture-based ground-truth motions, even before applying any inverse-kinematics postprocess (video\footnote{\href{http://y2u.be/lXd_7X-DkTA}{http://y2u.be/lXd\_7X-DkTA}}). 
Direct benefits of our approach could be to accelerate the creation of transition variations for large coverage, or even to entirely replace transition nodes in an animation graph. We further explore applications of this model in a animation super-resolution setting where we temporally decompress animations saved at 1 frame per second and show that the network is able to reconstruct motions that are hard to distinguish from un-compressed locomotion sequences.
\end{abstract}

%


%
%


\begin{teaserfigure}
     \includegraphics[width=1.0\textwidth]{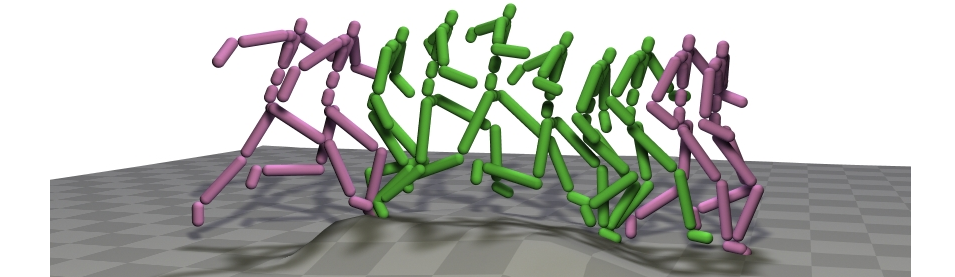}
  \caption{A transition animation (green) generated with our proposed system, from past and future contexts (magenta). Every fifth frame is shown.}
  \label{fig:teaser}
\end{teaserfigure}

\maketitle


\thispagestyle{empty} 

\section{Introduction}
Large games often require large animation graphs with a huge amount of unique animations in order to provide realistic and responsive movements. As the number of states grows in the graph, the number of required transitions may grow exponentially, while each of these transitions may require several variations to allow different conditions to be handled. In order to deal with continuous conditions or control signals, these variations of a single transition often need to respect additional constraints in order to allow interpolation, such as timing and phase constraints. All these elements make animation authoring an important but very tedious task. 

We explore in this research paper a data-driven approach based on deep Recurrent Neural Networks (RNN) to automatically generate transition clips from any character motion in order to reach a desired target, defined as the future desired state of the character. Such an approach aims at leveraging high quality data from Motion Capture (MOCAP) based animations, without needing labels, in order to greatly simplify the task of transition generation for games. While recent deep learning approaches have been proposed to improve motion prediction, or \textit{unconstrained continuation}, the problem of transition generation for games seems much less studied while being an important one. Our method builds on recent advances in deep learning for motion prediction such as the Encoder-Recurrent-Decoder (ERD) networks \cite{fragkiadaki2015recurrent} and the ResNet RNNs \cite{martinez2017human}, in which a recurrent network takes as input a number of frames that we call here the \textit{past context} and generates the next frames of animation. In our case, the architecture is designed specifically for transition generation by conditioning at every step of the generation on a target-relative representation called \textit{future context} that is evolving through time. The future context conditioning is done in a modified Long-Short-Term-Memory (LSTM) layer, responsible for modeling the temporal dynamics of motion, by adding extra weights and inputs for the computation of each gate and the internal cell values. We call such a network a Recurrent Transition Network (RTN). During training, we pick such past and future contexts directly from the data, without any labeling of gait, phase or other information. 

The hidden states' initialization of recurrent layers for a given sequence is often a subject absent from various work on motion prediction with RNNs. This is because default strategies, such as initializing these hidden states to zero-vectors or learning a common initial state as extra parameters, are used most of the time. It works in practice, especially when the RNN takes several frames as inputs to compute a good hidden state before starting the generation, but it unnecessarily complicates the task since the network has to learn to cope with this default initial hidden state by relying solely on the input for the first timesteps. We propose a simple way to initialize hidden states of recurrent layers with respect to any input sequence by having a small feed-forward network learning an inverse function that maps the first frame of the past context to the initial recurrent hidden state that minimize the generation error. This does not require any modification to the loss function, and improves the performance of the system.

We further explore additional constraint-conditioning by including local terrain information to the inputs of the network in the form of local height-maps based on current root (hips) location. In our experiments with longer transitions (2 seconds), we show that we have qualitative and quantitative gains over a terrain agnostic RTN.   

The recent focus of character kinematics modeling with deep recurrent neural networks has been mostly on motion prediction \cite{li2017auto, martinez2017human, ghosh2017learning} and online control \cite{holden2016deep, holden2017phase, zhang2018mode}, while animating transitions is still a core task for animators and could be greatly accelerated if not automated. Statistical latent variable models \cite{wang2008gaussian, lehrmann2014efficient} based on classical machine learning techniques for motion modeling and transition generation for single action types have been proposed, but the generality and scalability of deep neural approaches has yet to be employed specifically on this important task. 

Our core contribution for this paper can be summarized as novel, future-aware deep recurrent architecture specifically designed for transition generation, for which we motivate and quantify all the relevant design choices. It produces highly realistic and fluid character locomotion transitions while being generic, scalable and having a low memory footprint. This paves the way to greatly simplifying and accelerating transition authoring for games. 

\section{Related Work}
\subsection{Motion Control}
Our approach makes use of future targets to be reached in a fixed time in order to generate motion. This is related to the task of online control, where next frames or fragments of motion are generated with respect to higher-level signals, such as the desired trajectory, \cite{holden2016deep} or a footsteps plan, \cite{agrawal2016taskBasedLocomotion}. In these scenarios, the control signals often represent an under-specified description of the task, for which many possible executions may lead to the character completing it correctly. From that view, many control methods for locomotion are related to our work and relevant to mention.

Important work was done with motion graphs \cite{arikan2002interactive, lee2002interactive, kovar2008motion, beaudoin2008motion} in which a motion dataset is converted into a graph specifying the valid transitions for realistic movements. Walks through the graph then correspond to valid motions made from the visited nodes or edges depending on the method. Combined with a search algorithm \cite{safonova2007construction}, one can produce a transition generator that finds paths in the motion graph that reach the desired state in the desired time. These approaches, however, are limited to outputting motions from the dataset and often don't offer generalization capabilities comparable to deep learning methods. The graph and animations used must also be loaded in memory in order to output the resulting animation making the memory requirements scale with the dataset. 

\citet{chai2005performance} introduced a method that learns temporally-local low-dimensional linear models with PCA from nearest candidate poses to predict the next pose in the motion. Control signals here are low-dimensional vectors of marker positions or sparse accelerometer data in the case of \citet{tautges2011motion} and are mapped to full body animations. These require a preprocessed graph representation of the data available in memory and performing nearest neighbor search at runtime, and thus doesn't scale well with larger motion datasets.

Machine learning techniques and statistical models for motion control may offer advantages over graph-based methods, such as better generalization and lower memory requirements at runtime, at the cost of a computation-heavy learning phase. \textit{Maximum A Posteriori} (MAP) frameworks have been proposed to maximize the likelihood of generated motions with respect to environmental or user constraints, which can, similarly to our case, correspond to keyframes. \citet{chai2007constraint} propose a statistical dynamic model that can handle specific actions, with relatively close keyframes as constraints, while \citet{min2009interactive} build deformable motion models based on Principal Component Analysis (PCA) of the geometric variations and timing variations of a canonical motion cycle. These models are more compact than motion graphs, but specialized to certain motion cycles as they require run-time optimization of the MAP framework that doesn't scale well with a wide range of motions. 
Gaussian Process Latent Variable Models (GPLVM), which are framed as probabilistic, non-linear variants of PCA by \citet{lawrence2004gaussian, lawrence2005probabilistic} were also successfully applied to motion modeling and control \cite{grochow2004style, wang2008gaussian, ye2010synthesis, levine2012continuous}. \citet{min2012motion} also use Gaussian Processes to model transitions between morphable motion primitives. These methods often offer better generalization and allow to model uncertainty. However, they also suffer from limited scalability due to costly runtime computations and are often applied on individual action types, making the combination of actions seem very scripted (e.g. walk a step, then throw a punch). Our method is more generic and has a constant runtime, independent from the number of training samples.  

Methods based on neural networks have more recently shown impressive results for more generic kinematic control. Deep-learning based approaches allow to encode motion dynamics from large amounts of data into a small, fixed-sized network that is often fast to run. \citet{holden2016deep} present a deep-learning framework allowing animation synthesis from ground trajectories or desired end-effectors' positions, respecting a motion manifold learned by a convolutional auto-encoder \cite{holden2015learning}. Online control from user inputs was also tackled with deep learning work by \citet{holden2017phase} and Zhang, Starke et al. \shortcite{zhang2018mode} in which auto-regressive neural networks model the highly non-linear relations between a current character state, user inputs and the resulting next state. These methods rely on phase-dependent or mode-dependent weights to disambiguate the next possible pose. Our approach does not require similar explicit disambiguation techniques since our network can model high-level temporal dynamics through its recurrent layer and further uses future context information as a guiding signal.   

The motion control problem has also been addressed through reinforcement learning (RL) methods. Some previous methods made use of animation clips \cite{lee2006precomputing, treuille2007near} or motion states \cite{lee2010motion} taken directly from the data as outputs of the RL algorithm, thus requiring large amounts of memory for large datasets. \citet{levine2012continuous} learn a control policy operating in the low-dimensional latent space learned from their GPLVM variant for which computations scale linearly with the number of samples in the data, thus limiting the system to model single action classes. In the domain of physically-based locomotion, RL techniques have also been applied, such as fitted value iteration \cite{coros2009robust}, for which actions correspond to pre-optimized proportional-derivative (PD) controllers for locomotion \cite{yin2007simbicon}. These techniques all rely on value functions that have discrete domains and rely on interpolation during runtime to handle continuous states, limiting to some extent the expressiveness of the systems. 

The combination of deep learning with advances in reinforcement learning gave rise to deep reinforcement learning methods that addressed many limitations of these previous approaches. \citet{peng2017deeploco} apply a hierarchical actor-critic algorithm that outputs desired joint angles for PD-controllers. Their approach is applied on a simplified skeleton and does not express human-like quality of movement despite their style constraints. Imitation-learning based RL approaches \citep{ho2016generative, baram2016model} try to address this with adversarial learning, while \citet{peng2018deepmimic} tackle the problem by simply penalizing distance of a generated state from a reference state and proposing new training strategies, such a reference state initialization. Actions as animation clips, or control fragments \cite{liu2017learning} can also be used in an RL framework with Q-learning to drive physically-based characters. These methods show impressive results for characters having physical interactions with the world, while still being limited to specific skills or short cyclic motions. We operate in our case in the kinematics domain and train on significantly more diversified motions.

\subsection{Motion Prediction}
We limit here the definition of motion prediction to predicting the motion continuation given single or multiple frames of animation. This unrestricted formulation alone may have limited concrete applications for games, where movements are constrained by control or targets signals, but it is nonetheless very relevant to learn a powerful dynamics' model in such a way for transition generation.
Neural networks have shown over the years to excel in such representation learning. Early work from \citet{taylor2007modeling} using Conditional Restricted Boltzmann Machines showed promising results on motion generation by sampling at each timestep the next frame of motion conditioned on the current hidden state and $n$ previous frames. More recently, many RNN-based approaches have been proposed for motion prediction from a past-context of several frames, motivated by the representational power of RNNs for temporal dynamics. \citet{fragkiadaki2015recurrent} propose to separate spatial encoding and decoding from the temporal dependencies modeling with the Encoder-Recurrent-Decoder (ERD) networks, while \citet{jain2016structural} apply structural RNNs to model human motion sequences represented as spatio-temporal graphs. \citet{martinez2017human} propose to simplify these networks with a single-layer Residual Recurrent Network predicting offsets from the current pose in order to reduce jumps on the first predicted frame. In this work, we use an architecture closer to ERD networks and use different separate encoders for the different inputs, in order the allow for a smaller recurrent layer focused on conditioning the temporal dynamics on various signals. \citet{li2017auto} and \citet{ghosh2017learning} investigate ways to prevent divergence on collapsing to the average pose for long-term predictions with RNNs. In our case, these problems are naturally mitigated by the added constraints of the target state to obtain and the reduced length of the generation. We also use a fixed-probability teacher-forcing scheme that empirically showed improvements over other teacher-forcing strategies, as described in Section \ref{sec:teacher}.    

\subsection{Transition generation}
Work applied directly to transition generation is sparser, especially in the deep learning literature. However, important work with probabilistic models of human motion have been used for transition generation, or gap-filling of animation. These include the MAP optimizers of \citet{chai2007constraint} and \citet{min2009interactive}, the Gaussian process dynamical models from \citet{wang2008gaussian} and Markov models with dynamic auto-regressive forests from \cite{lehrmann2014efficient}. All these present specific models for given action and actors, making combinations of actions hard or unnatural, while we leverage a large heterogeneous dataset along with the scalability of deep learning methods to train a single, generic model trained with unlabeled and unsegmented motion sequences.

\section{Data formatting}
\subsection{Dataset}
The data for this experiment was captured in a MOCAP studio using a Vicon system. It contains a lot of unstructured motion data summarized in Table \ref{table:dataset}, all re-targeted to a common skeleton from which we use $K=22$ bones, and downsampled at a rate of 30 frames per second.
\begin{table}[h]
\caption{Dataset motion categories}
\begin{sc}
\begin{center}
\begin{tabular}{lrr}
Motion &Frames (at 30 fps) &Minutes
\\ \hline
Flat Locomotion     & 240 776 & 133\\
Terrain Locomotion  & 113 020 &  63\\
Dance               &  38 916 &  22\\
Others              & 115 716 &  64\\
Total               &508 428 &  282\\
\end{tabular}
\end{center}
\end{sc}
\label{table:dataset}
\end{table}
The \begin{sc}Others\end{sc} category contains fighting movements, falling down and recovery motions, as well as sports-like movements and other miscellaneous motions. No labeling was done on the sequences, and the 5 performers were not professional actors. The locomotion components of the dataset are similar to the publicly available terrain locomotion data from \citet{holden2017phase}. In all of our experiments except for those detailed in Section \ref{section:superres}, we used only the terrain locomotion data, corresponding to approximately one hour of MOCAP. The dataset contains long, uninterrupted sequences of motions that we split into $N$ overlapping sub-sequences $\bY_n$, $n \in \{0, ..., N-1\}$ that have a length $L$ determined by our desired transition length. As we use a past context of 10 frames in all of our experiments, and 2 frames are needed to produce our target state, we have a necessary $L \geq 10 + P + 2$, where $P$ is our desired transition length. In practice however, as we want to qualitatively assess the continuity of the movement after the transition, we keep 10 post-transition frames in our windows. Therefore, unless specified otherwise, we use $P=30$ in our experiments, as discussions with domain experts led us to believe that such a length covers a vast majority of transitions for many games. 

\subsection{Input representations and processing}
\subsubsection*{Input sequences}
Similarly to \cite{holden2016deep}, we work with positional information. The raw data consists of sequences $\bY_n = \{\by_0, ..., \by_{L-1} \}_n$ of vectors $\by_t$ of global 3D positions. The dimensionality of all positional or velocity vectors in this work is of $D = 3 \times K$ where $K=22$ is the number of bones used. Our preprocessing of the positional vectors $\by_t$ consists of three steps. 

First, we store in $\bv_t \in \mathbb{R}^3$ the 3D global velocity of the root joint (\textit{hips}) at each frame $t$, computed from the instant offset from the previous frame. Hence, $\bv_t = \br_t - \br_{t-1}$, where $\br_t$ is the global root position extracted from $\by_t$.

Secondly, we remove from every marker the global position of the root joint in order to get root-relative positions. We then replace the now null positional root information in the vector by its instant global velocity. This yields a new vector $\tilde{\bx}_t = [\bv_t, \, \bj^1_t, ...,\, \bj_t^{K-1}]^\mathrm{T}$ where $\bj^k_t$ is the 3D root-relative position of joint $k$, with indexes starting at $0$ for the root. 

The third and last step is the normalization scheme commonly used in deep learning in order provide the network data samples following zero-centered Gaussian distribution with an identity covariance matrix. In the rest of the paper, we will use the term \textit{normalize} to represent this \textit{standardization} operation of subtracting from a data sample $\bz$ its mean $\bm{\mu}_z$ computed over the training set of the data, and dividing the result by the standard deviation $\bm{\sigma}_z$ over the training data. We therefore have our processed input vector $\bx_t = (\tilde{\bx}_t - \bm{\mu_x}) / \bm{\sigma_x}$.

Our preprocessing function is reversible, as long as $\bm{\mu_x}$ and $\bm{\sigma_x}$ are stored, as well as a global offset information for a given sequence $\bY_n$, such as the global position of the root on the first frame, in order to place it back correctly in the world frame.  We purposely defined a separate notation for the un-processed positional frames $\by_t$ as they will be used when handling local terrain information.

\subsubsection*{Future context}
In order to generate transitions, the system makes use of information relative to the desired target state. We call this the future context. It is used as conditioning information at every timestep and consists of the concatenation of two different vectors. 

The first one is the target vector $\bt \in \mathbb{R}^{2*D}$ which is simply the processed target pose we want the character to end up in concatenated with the normalized velocity of all joints on that frame. The normalization step of the velocity is done with statistics computed over the preprocessed training set. The target vector is constant throughout the generation process. 

The other vector included as future context is the global offset vector $\bo_t \in \mathbb{R}^D$ which is composed of the euclidean distances of each joints from the target pose in global space (unpreprocessed). As the transition is generated, these global offsets evolve with respect to the last generated frame and the global target positions. The vector $\bo_t$ is also normalized with the training set statistics.   

\subsubsection*{Terrain}
In most of our experiments, we also make use of local terrain information as an additional guiding signal for better reconstructions. Similarly to \cite{peng2017deeploco, peng2018deepmimic}, we use a local heightmap relative to the $(x,z)$ root position. In our case, we sample at each timestep a 13 $\times$ 13 grid spanning 2.06 $\times$ 2.06 meters centered on the character. We then convert this heightmap into a grid of $y$-offsets from the root joint. This is to allow a local terrain representation that is invariant to the general elevation of the area. We then normalize this grid on statistics computed on the training set.
\begin{figure}[h]
\begin{center}
\centerline{\includegraphics[width=0.3\textwidth]{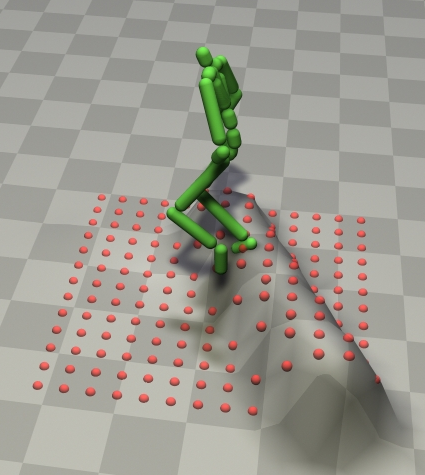}}
  \caption{Visual depiction of a local terrain patch.}
  \label{fig:local_patch}
\end{center}
\end{figure}
We use the 169-dimensional flattened vector $\bp_t$ of this grid as our representation for the local patch of terrain. A visual depiction of the unprocessed information contained in $\bp_t$ is shown in Figure \ref{fig:local_patch}.

\subsection{Data augmentation}
\subsubsection*{Terrain}
As the motion dataset does not contain any obstacle or terrain information from which to extract the local patches $\bp_t$, we first need to augment our data with coherent terrain information with respect to the motion. We follow the terrain fitting procedure proposed by \citet{holden2017phase}, which consists of fitting and editing several heightmaps $\bH_{l}, l\in \{0,...,4\}$ with respect to each sequence $\bY_n$ that are large enough to cover the range of the motion on the transverse plane. It is done by first evaluating a cost function that penalizes foot traversal and floating based on feet contacts and heights at each timestep. The contact information is extracted automatically from feet velocities. This fitting step is done for all sequences and heightmaps. A refinement steps then applies to the best scoring heightmaps a 2D radial basis function (RBF) on the feet $y$-offsets from the terrain on contact frames, which is evaluated and added to the terrain as corrections. We refer readers to \cite{holden2017phase} for more details, as we use the same cost function, and the same heightmap dataset.

Our method differs in that we use a logistic kernel for the RBF instead of a linear one, while also further smoothing the resulting heightmaps with a 2D Gaussian filter around the contact points on the condition that it doesn't decrease the overall terrain score for the sequence. Our filter $\bF$ is a 33 $\times$ 33 grid spanning 1.3 $\times$ 1.3 meters with values following an isotropic 2D Gaussian with a standard deviation of 5 (20 centimeters). It is re-scaled so that its maximum value is equal to one. We then correct the values of the heightmap $\bH$ over a sub-grid with the same shape as $\bF$ around the point of contact $(x, z)$ with the $\bF$-weighted $y$-offset $\Delta_y$ of the contact foot from the ground:
\begin{equation}
    \bH_{x+i-16, z+j-16} = \bH_{x+i-16, z+j-16} +  \Delta_y  \bF_{i,j}
\end{equation}
This is done for every contact point. These modifications to the algorithm are made to provide the system with the most realistic-looking terrain during training, to reflect plausible locomotion conditions. We keep and refine the top-5 terrain heightmaps for every sub-sequence $\bY_n$ in the dataset, and pick one randomly at each training iteration as our data augmentation.

\subsubsection*{Random orientation}
Our data preprocessing of the global positions makes the motion modeling translation invariant since we work with root-relative positions for all joints, and global velocity for the root. We enforce horizontal orientation invariance as well by randomly rotating each motion sequence $\bY_n$ seen during training around a unit \textit{up}-vector emerging from the center of the terrain. The angle of rotation is drawn uniformly in $[-\pi, \pi]$. The rotation is performed on the global positions, before any pre-processing of the sequence. We apply the same rotation on the terrain heightmap around its center in order to extract local patches $\bp_t$ consistent with the new global $(x,y)$ coordinates of the positions.

\section{Recurrent Transition Network}\label{section:architecture}
\subsection{System overview}\label{section:overview}
\begin{figure}[h]
\begin{center}
\centerline{\includegraphics[width=0.36\textwidth]{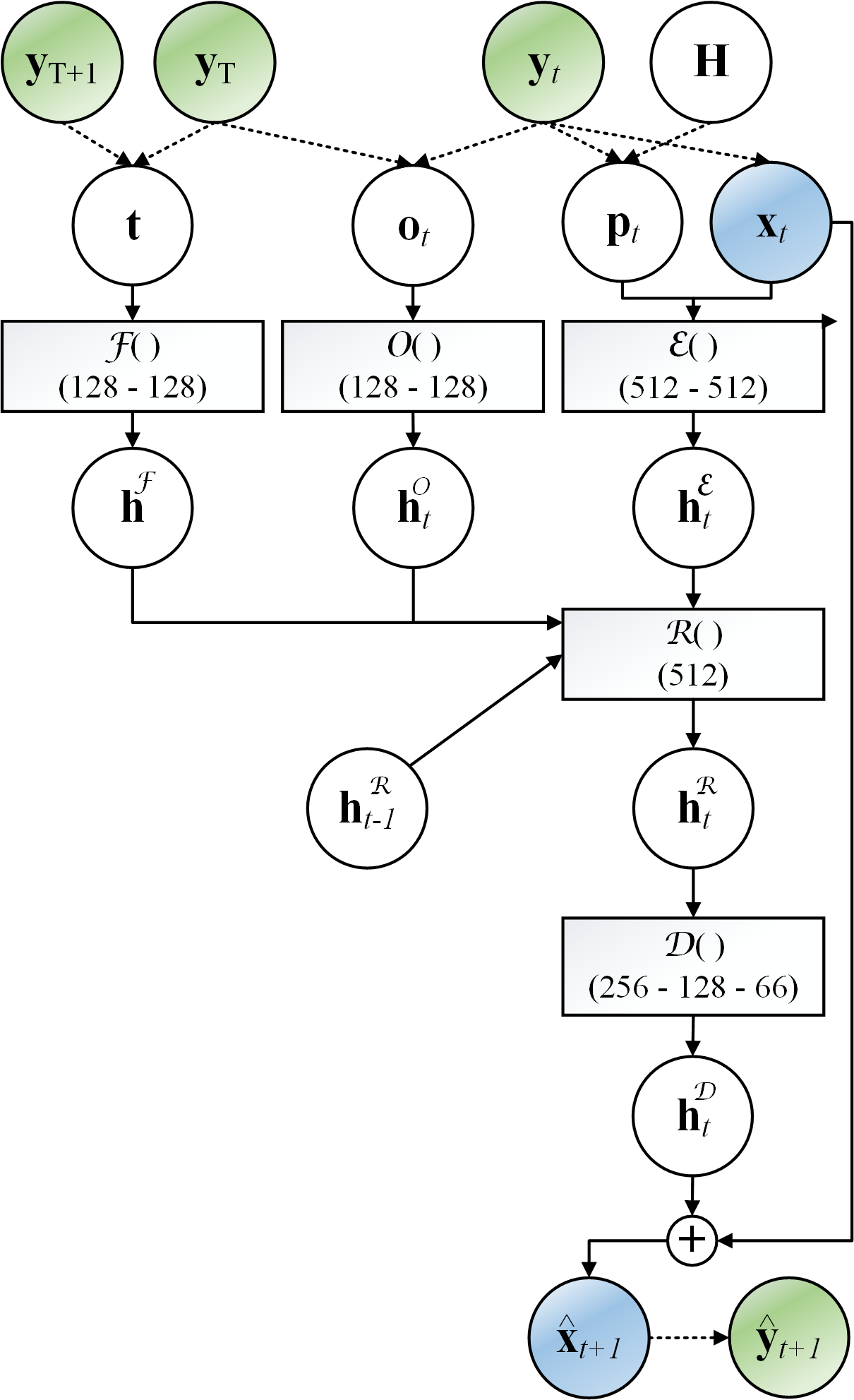}}
  \caption{Graphical overview of the proposed architecture for a single step of generation. Numbers in parenthesis correspond to the number of hidden units used in all of our experiments. Dashed lines represent operations happening outside the neural network. Green circles represent un-preprocessed, global positional data, while blue circle represent preprocessed data.}
  \label{fig:overview}
\end{center}
\end{figure}

An overview of our proposed system can be seen in Figure \ref{fig:overview}, depicted for a single timestep. It takes as inputs a preprocessed positional vector $\bx_t$ retrieved from the unprocessed corresponding global positional vector $\by_t$ and a local terrain patch representation $\bp_t$ retrieved from a heightmap $\bH$ with respect to $\by_t$. It uses as conditioning information a normalized target vector $\bt$ computed from the global positions on the target frame $\by_T$ and the next one $\by_{T+1}$, and a global-offset vector $\bo_t$ retrieved from $\by_t$ and $\by_T$. 
The RTN is composed of several, specialized sub-networks all trained jointly by gradient descent. Similarly to ERD networks, a frame encoder $\mathcal{E}()$ and frame decoder $\mathcal{D}()$ are used as spatial encoder/decoders, while the temporal dynamics are modeled in a modified LSTM layer $\mathcal{R}()$. Two smaller encoders are also used for encoding parts of the future context. The target encoder $\mathcal{F}()$ encodes the desired normalized target state to reach, once per sequence, while the global-offset encoder $\mathcal{O}()$ is used to encode at every step the current global positional offset from the target. An additional, optional hidden state initializer small sub-network $\mathcal{H}()$ is used in our case once per sequence in order to initialize the hidden state and cell of the LSTM layer appropriately for every sequence. All the operations of these sub-networks are detailed in the rest of this Section.

\subsection{Frame encoder}
At each timestep, the preprocessed current character configuration $\textbf{x}_t$ and local patch $\bp_t$, are first transformed into a new hidden representation $\bh^{\mathcal{E}}_t$ by the frame encoder $\mathcal{E}()$, which consists of a Multi-Layer Perceptron (MLP) with two hidden layers:
\begin{equation}
    \bh^{\mathcal{E}}_t = \mathcal{E}(\bx_t, \bp_t) = \phi \big(\bW^{(2)}_{\mathcal{E}} \phi(\bW^{(1)}_{\mathcal{E}} 
    {\begin{bmatrix}
    \bx_t \\
    \bp_t\\
    \end{bmatrix}}
    + \bb^{(1)}_{\mathcal{E}})+ \bb^{(2)}_{\mathcal{E}}\big)
\end{equation}
where $\bW^{(l)}_{\mathcal{E}}$ and $\bb^{(l)}_{\mathcal{E}}$ are the weight matrices and bias vectors of the $l^\mathrm{th}$ layer, and $\phi$ is the Leaky Rectified Linear Unit (LReLU) activation function \cite{maas2013rectifier}. The frame encoder has 512 units in both its layers. In experiments without terrain-awareness, the concatenation with $\bp_t$ is absent and $\bx_t$ only is received as a lower-dimensionality input.

\subsection{Future context encoders}
The future, normalized target $\bt$ encoder $\mathcal{F}()$ and the global offset $\bo_t$ encoder $\mathcal{O}()$ are both similar MLPs to the frame encoder, with two 128-dimensional hidden layers each.
\begin{align}
    \bh^{\mathcal{F}} = \mathcal{F}(\bt) &= \phi \big(\bW^{(2)}_{\mathcal{F}} \phi(\bW^{(1)}_{\mathcal{F}} \bt + \bb^{(1)}_{\mathcal{F}})+ \bb^{(2)}_{\mathcal{F}}\big)\\
    \bh^{\mathcal{O}}_t = \mathcal{O}(\bo_t) &= \phi \big(\bW^{(2)}_{\mathcal{O}} \phi(\bW^{(1)}_{\mathcal{O}} \bo_t + \bb^{(1)}_{\mathcal{O}})+ \bb^{(2)}_{\mathcal{O}}\big)
\end{align}
The future target $\bt$ is encoded once and is constant for every timestep of a given sequence.

\subsection{Recurrent generator}
The recurrent generator $\mathcal{R}$ is responsible for the temporal dynamics modeling and the conditioning on the future context. It is a single 512-dimensional LSTM layer that uses the concatenation $\bh_t^{\mathcal{F,O}}$ of $\bh^{\mathcal{F}}$ and $\bh^{\mathcal{O}}_t$ as future-conditioning information along with added corresponding parameters. The LSTM equations are as follow :  

\begin{gather}
\bi_{t} = \alpha(\bW^{(i)} \bh^{\mathcal{E}}_t + \bU^{(i)} \bh^\mathcal{R}_{t-1} + \bC^{(i)} \bh_t^{\mathcal{F,O}} + \textbf{b}^{(i)})\\
\bo_{t} = \alpha(\bW^{(o)} \bh^{\mathcal{E}}_t + \bU^{(o)} \bh^\mathcal{R}_{t-1} + \bC^{(o)} \bh_t^{\mathcal{F,O}} + \textbf{b}^{(o)})\\
\bff_{t} = \alpha(\bW^{(f)} \bh^{\mathcal{E}}_t + \bU^{(f)} \bh^\mathcal{R}_{t-1} + \bC^{(f)} \bh_t^{\mathcal{F,O}} + \textbf{b}^{(f)})\\
\hat{\bc}_{t} = \bW^{(c)} \bh^{\mathcal{E}}_t + \bW^{(c)} \bh^\mathcal{R}_{t-1} + \bC^{(c)} \bh_t^{\mathcal{F,O}} + \bb^{(c)}\\
\bc_{t} = \bff_t \odot \bc_{t-1} + \bi_t \odot \tau(\hat{\bc}_{t})\\
\mathcal{R}(\bh^{\mathcal{E}}_t, \bh^\mathcal{R}_{t-1}, \bc_{t}, \bh^{\mathcal{O}}_t, \bh^{\mathcal{F}}) = \bo_{t+1} \odot \tau(\bc_{t})\\
\bh^\mathcal{R}_{t} = \mathcal{R}(\bh^{\mathcal{E}}_t, \bh^\mathcal{R}_{t-1}, \bc_{t}, \bh^{\mathcal{O}}_t, \bh^{\mathcal{F}})
\end{gather}
where $\bW^{(\{i,o,f,c\})}$, $\bU^{(\{i,o,f,c\})}$, $\bC^{(\{i,o,f,c\})}$ are feed-forward, recurrent, and conditioning weight matrices respectively, $\bb^{(\{i,o,f,c\})}$ are bias vectors, and $\odot$ is an element-wise multiplication. The $\alpha$ and $\tau$ functions are the commonly used \textit{sigmoid} and \textit{tanh} non-linearities of LSTM networks.

\subsection{Frame decoder}
Each of the generated output $\bh^\mathcal{R}_t$ of the LSTM generator is passed to the frame decoder $\mathcal{D}$, which is another fully-connected feed-forward neural network. It has two hidden layers of 256 and 128 LReLU-activated neurons, and a D-dimensional linear output layer.
\begin{equation}
    \bh^{\mathcal{D}}_{t} =\mathcal{D}(\bh^\mathcal{R}_{t}) = \bW^{(3)}_{\mathcal{D}} \phi\big(\bW^{(2)}_{\mathcal{D}} \phi(\bW^{(1)}_{\mathcal{D}} \bh^\mathcal{R}_{t} + \bb^{(1)}_{\mathcal{D}}) + \bb^{(2)}_{\mathcal{D}}\big) + \bb^{(3)}_{\mathcal{D}}
\end{equation}
We use a Res-Net LSTM, which outputs an offset from the current frame $\bx_t$, as suggested by \citet{martinez2017human} in order to reduce the gap between the input seed frames and the beginning of the transition. The final prediction $\hat{\bx}_{t+1}$ is therefore obtained with:
\begin{equation}
\hat{\bx}_{t+1} = \bx_{t} + \bh^{\mathcal{D}}_{t}
\end{equation}

From $\hat{\bx}_{t+1}$, we can retrieve $\hat{\by}_{t+1}$ by applying the inverse of our preprocessing function, which consists of de-normalizing the vector, retrieving the global root positions by cumulatively adding the velocities to the stored first global frame of the sequence, and then adding those root positions to the other joints.

\subsection{Hidden state initializer}
\begin{figure}[h]
\begin{center}
\centerline{\includegraphics[width=0.32\textwidth]{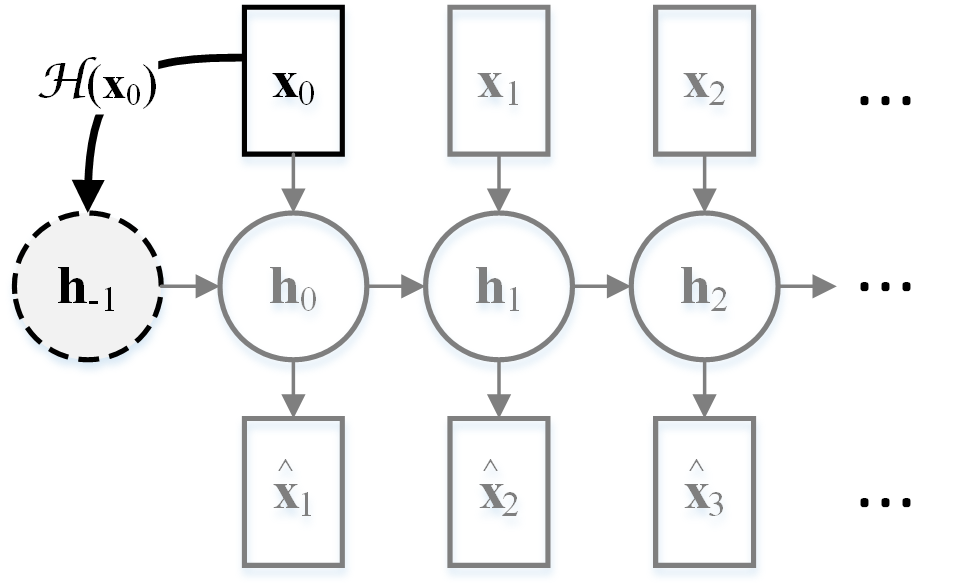}}
  \caption{Graphical overview of the proposed hidden state initialization application. The function $\mathcal{H}()$ maps $\bx0$ to the initial hidden state $\bh_{-1}$ that will minimize the loss.}
  \label{fig:hi}
\end{center}
\end{figure}

It is often the case that for continuous prediction, such as motion generation, initialization details for the hidden state\footnote{We include in $\bh_{-1}$ here both the \textit{output} and \textit{cell} states of an LSTM layer at $t=-1$.} $\bh_{-1}$ of any recurrent layer is left unspecified (e.g. \cite{ghosh2017learning, martinez2017human}).  One can assume that in these cases, the initialization procedure can very well be the default initialization scheme defined by the chosen deep learning library. The most common strategy in this case is to initialize those states to zero vectors, as done by \citet{li2017auto}, or to treat this initial state as additional parameters for the network to learn. With such strategies however, the network must learn to handle this special initial vector common to all samples, and it may then become necessary to provide many input frames to reach a good recurrent internal state encoding properly the motion to complete. We instead propose a simple, yet principled way to initialize $\bh_{-1}$ with respect to a given sequence. Our method consists of learning an inverse function that predicts $\bh_{-1}$ given the first frame of the input sequence $\bx_0$. We learn this with an additional, single hidden layer MLP:
\begin{equation}
\bh_{-1} = \mathcal{H}(\bx_0) = \bW^{(2)}_{\mathcal{H}} \phi(\bW^{(1)}_{\mathcal{H}} \bx_0 + \bb^{(1)}_{\mathcal{H}})+ \bb^{(2)}_{\mathcal{H}}
\end{equation}
For which the output constitutes the initial hidden sate. Figure \ref{fig:hi} shows where this sub-network is used in practice.
 This method allows for the recurrent hidden states to be initialized differently for any given input into a good region of the hidden space. In other words, at time $t=0$ the recurrent layer can use a valid input ($\bx_0$) \textit{and} a valid previous hidden state ($\bh_{-1}$) that should reduce reconstruction error of the next frames. We train the additional parameters of this sub-network jointly with the rest of the system, without any modification to the loss function, as opposed to \citet{mohajerin2017state} who design a more complex hidden state initialization procedure that requires a separate loss function and a sub-sequence of timesteps as inputs for the state initializer. Our method could work with past contexts of a single frame, and improved the generalization capabilities of the system in all our experiments over commonly used techniques. 

\section{Postprocessing}
Our network produces realistic transitions without the need for much postprocessing. In every sample showed, no IK pass is done on top of the generated sequence to prevent feet artifacts. The only postprocessing applied is a correction of the gap between the last generated frame and the target position. We refer to this naive postprocessing as \textit{target blend}.

\subsection{Target blend}
As our system generates a transition to desired a state, it uses this additional target information to shape its prediction as opposed to only predicting the next poses. We can also make use of this additional information to postprocess the generated animation sequences. The network always predicts an additional character-state on the target frame. This allows us to quantify with an offset vector $\be = \by_{T} - \hat{\by}_{T}$ the distance of the true positional target $\by_{T}$ from the estimated targeted positions $\hat{\by}_{T}$. With that quantity we can then correct a generated transition frame $\hat{\by}_t$ with 
\begin{align}
    \tilde{\by}_t &= \hat{\by}_t + \omega \be,\\
    \omega &= 1 - \frac{(T-t)}{d}
\end{align}
where $\omega$ is an increasing weight that depends on the current frame index $t$, the target frame index T and the blend duration $d$. The target blend is computed only on frames inside the blend duration. The target blend can smoothly prevent a small jump from the last generated frame and the target position, but will result in artifacts such as foot-skate when a network such as a standard RNN fails to produce a last transition frame that is sufficiently close to the target.

\section{Training}
We will describe here the training procedure details for the RTN. 
\subsection{Loss}Throughout the experiments, our loss is defined as the Mean Squared Error (MSE) on the reconstructed transition $\hat{\bX} = \{\hat{\bx}_{s}, ..., \hat{\bx}_{T}\}$ :
\begin{equation}
    MSE(\bX, \hat{\bX}) = \frac{1}{T-s} \sum_{t=s}^{T} ||\hat{\bx}_{t} - \bx_{t}||^2
\end{equation}
where $s$ is the time index of the first transition frame and $T$ is the time index of the target position to reach. The RTN predicts this target in addition to the whole transition as this can allow to quantify the offset in prediction for that target and perform target-blending, as described above. Therefore, when aiming at generating transitions of $P$ frames, the RTN is actually trained to predict $P+1$ frames.   

\subsection{Probabilistic Teacher Forcing}\label{sec:teacher}
One core design choice when training RNNs for prediction is on the amount of teacher forcing that is performed during training. Teacher forcing consists of providing to the RNN the ground-truth past frame during training instead of its own previous prediction \cite{williams1989learning}. This can accelerate learning as the generator stays in the realm of plausible motions during training and won't diverge into unlikely trajectories and try to correct for them with exaggerated gradient updates. However, this may also limit the capacity of the network to compensate for its accumulation of errors when deployed or at test time, making it diverge from plausible trajectories. Scheduled sampling \cite{bengio2015scheduled} was proposed in order to have a good balance between training time and test performance. It consists of choosing at each timestep whether to pick the previous true frame $\bx_{t-1}$ with a probability of $p$ or the  previous prediction $\hat{\bx}_{t-1}$ with a probability of $1-p$ and decreasing $p$ during training. \citet{li2017auto} also investigated this aspect, applied on realistic motion prediction and propose to have a constant-length window during which $\bx_{t-1}$ is used and another fixed-length window in which $\hat{\bx}_{t-1}$ is used and alternating these windows during training. They show that this Auto-Conditioned LSTM (AcLSTM) for long-term predictions can produce remarkably stable and plausible trajectories at test-time, even though it may not get the lowest reconstruction error. In our case, we use a fixed-probability version scheduled sampling, which shows improved performance over the vanilla training scheme, teacher forcing, scheduled teacher forcing and AcLSTM both quantitatively and qualitatively.

\subsection{Hyperparameters}
We perform stochastic gradient descent training with minibatches of size 32, using the recently proposed AMSGrad optimization procedure \cite{reddi2018convergence}. We set the probability $p$ to $0.2$ for our fixed ground-truth sampling strategy. Our learning rate is set 0.0005, while we use the common default values of ADAM \cite{kingma2014adam} of $\beta_1$ and $\beta_2$ for the AMSGrad optimizer. We do not use regularization schemes, such as the L2 regularizer or dropout, as it showed no improvement, or decreased performance in our empirical study. For the case of one-second transitions, with the terrain locomotion dataset, the data is processed into 5619 overlapping windows of 50 frames, of which 10 are used as past context, 30 are used as transition ground truth and the remaining 10 are used to extract the target and visualize the quality of the continuation of the movement after the transition. We extract all windows of motion performed by a certain actor to create our validation set. We run the training for 200 epochs in all experiments and show results on the best performing network iteration on the validation set. Details of the data splitting for other setups such as 2-seconds transitions or when using all the data are found in Appendix \ref{app:A}.

\section{Results}
We discuss here the different experiments and results obtained with the RTN, and assess many of the design choices made for transition generation. All samples shown here present results from the validation set, with motions from an actor not seen during training. We refer the reader to the additional supplementary video for a better view of the samples, and many additional results.

\subsection{Transition Reconstruction}
We first train our network to generate transitions of 30 frames on the rough terrain locomotion dataset, consisting of approximately one hour of motion, from which we extract all sequences from one of the five subjects as validation data. This amounts to 20\% of the data, as all actors are similarly represented. Figure \ref{fig:teaser} shows a transition during a forward leap. In general, such locomotion is very well handled by the network, which does not suffer from diverging trajectories or collapsing to the average pose. In most cases, without any post-processing of the resulting animation, it is hard for external viewers to determine if a given transitions was generated or coming from the data. This phenomenon is amplified when correcting the generated transitions with the target blend procedure detailed above. The system successfully models walk and run cycles, turning motions, changes in velocities and stances, all with a wide range of gaits, without using any phase or contact information. The network uses under 15MB of memory. The future context representation and conditioning schemes allow the network to reach the desired targets in time, with a minimal offset easily removed by the target blend.

As our system has a core architecture similar to the Encoder-Recurrent-Decoder (ERD) proposed by \citet{fragkiadaki2015recurrent}, we compare our approach to a baseline consisting of an ERD network with the same number of units for the corresponding components. We augment this ERD network with our future-context conditioning LSTM layer (\begin{sc}F-ERD\end{sc}) as a better suited baseline for the given task, since future agnostic networks cannot infer targets from past context. We augment in a similar fashion a single-layer ResNet-LSTM (\begin{sc}F-ResLSTM\end{sc}) inspired by the architecture proposed by \citet{martinez2017human} as another baseline.
Finally, since we perform the task of transition generation, we have access to the target position so we can compare our method with a more naive interpolation strategy (\begin{sc}INT\end{sc}). In this case, we use the positional and angular information from the last frame of the past context and from the target frame. From those, we perform linear interpolation on the local positions and spherical interpolation on the local quaternions of the bone hierarchy, and retrieve the resulting interpolated global positions. Table \ref{table:res1} summarizes these results. The MSE is calculated in the preprocessed space, so we also provide a measure of Average Centimeter Offset, consisting of the absolute centimeter offset from the ground truth of the generated transition, averaged over all degrees of freedom and over time.

\begin{table}[h]
\caption{Comparison of methods on the MSE on the validation set.}
\begin{sc}
\begin{center}
\begin{tabular}{lcc}
Architecture &MSE &ACO
\\ \hline
INT      &0.210 &6.726\\
F-ResLSTM&0.144 &7.709\\
F-ERD    &0.092 &4.770\\
RTN      &\textbf{0.087} &\textbf{4.751}\\
\end{tabular}
\end{center}
\end{sc}
\label{table:res1}
\end{table}

\begin{figure}[h]
\begin{center}
\centerline{\includegraphics[width=0.49\textwidth]{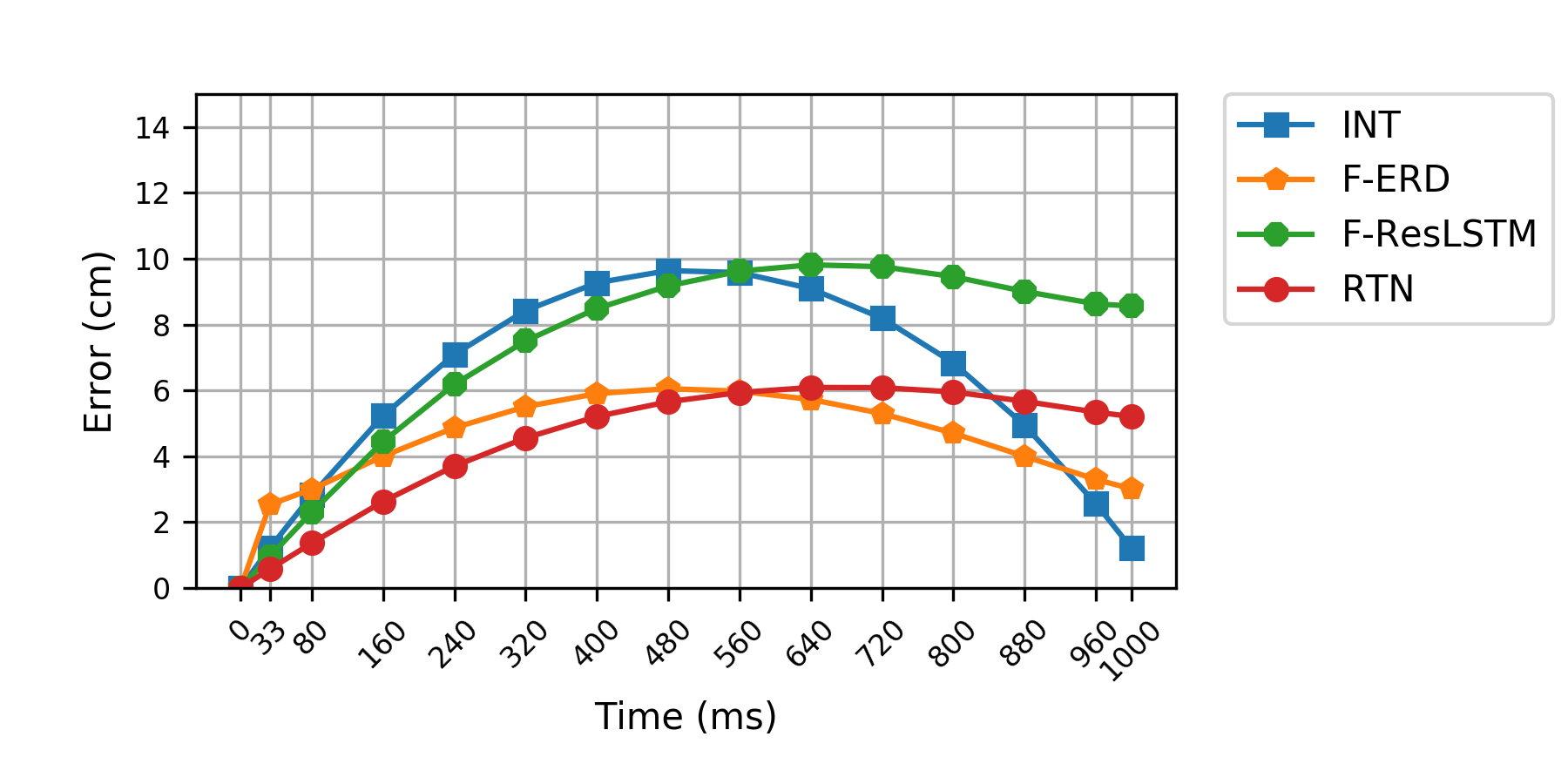}}
  \caption{Comparison of absolute centimeter offsets at fixed times for different methods.}
  \label{fig:perf1}
\end{center}
\end{figure}

We also present in Figure \ref{fig:perf1} a refined comparison of these models on the average absolute centimeter offset per degree of freedom at fixed times. This gives summary profiles of performance over time for the methods.
Unsurprisingly, naive interpolation produces good results on the firsts and lasts generated frames, but diverges from the ground truth when far from the end and start points. The \begin{sc}F-ResLSTM\end{sc} struggles to stay close to the true transitions even though it shows good short-term motion continuation. This may be due to the fact that mixing future-context conditioning to the spatial and temporal encoding tasks of the LSTM becomes too complex in a single layer. The \begin{sc}F-ERD\end{sc} network, which is the most similar architecture to the complete RTN, shows an interesting trade-off. It produces a gap on the first frame of the generation, degrading the motion continuation, but this seems to facilitate the task of reaching the target, perhaps because the first-frame offset starts the generation closer to the target frame. These offsets from \begin{sc}F-ERD\end{sc} could be smoothed with target blending as well, but would require two passes of postprocessing for the start and end of generation.

\subsection{Impact of terrain awareness}
\begin{figure}[h]
\begin{center}
\centerline{\includegraphics[width=0.45\textwidth]{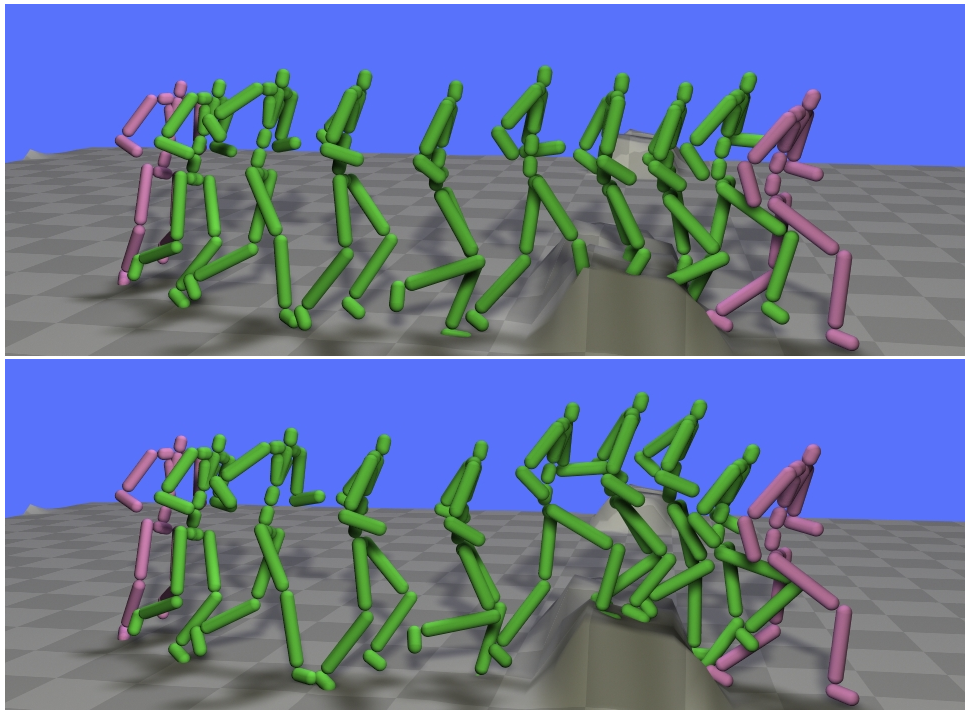}}
  \caption{Example of a generated 2-second transition with an obstacle. TOP: The terrain-unaware version of the RTN. BOTTOM: The terran-aware version of the RTN.  Every eighth frame is shown.}
  \label{fig:obstacle}
\end{center}
\end{figure}
The terrain information is used in our system to mitigate feet artifacts such as traversal, floating, or foot-skate. This information is added to the input representation to help the RTN learn relations between feet and terrain thus reducing the loss and producing better transitions. Such reduction of artifacts could reduce the need for an inverse-kinematics pass as post-process, or at least reduce the modifications it entails. However, as our experiments show, transitions of 30 frames seem too short for such added information to improve performance. Indeed, on clips with such lengths, it is hard to find samples where the character goes through an obstacle for which terrain information would be necessary to prevent. With transitions of 60 frames however, although the overall performance of the RTN decreases, the addition of terrain information to the inputs improves the performance of the system, as shown in Table \ref{table:res2}, and is easily noticeable on key samples where obstacles must be acknowledged for plausible transitions. An example of such a sample is shown in Figure \ref{fig:obstacle}.

\begin{table}
\begin{sc}
    \caption{Impact of terrain awareness on the MSE for transitions of different lengths.}
    \begin{tabular}{lrr}
        &\multicolumn{2}{c}{Transition length}
        \\\cmidrule(r){2-3}  
        &30 frames &60 frames\\\midrule
        Terrain-unaware      & \textbf{0.086} & 0.274\\
        Terrain-aware        & 0.087 & \textbf{0.268}\\
    \end{tabular}
    \label{table:res2}
\end{sc}
\end{table} 

\subsection{Ablation study}
In order to assess the benefit of the different modifications we bring to the ERD network, an ablation study is done in which we remove or modify individual components or sub-methods to the RTN. Table \ref{table:ablation} shows the quantitative effects of such modifications.
\begin{table}[h]
\caption{Ablation study on the RTN network.}
\begin{sc}
\begin{center}
\begin{tabular}{lc}
Variant &MSE
\\ \hline
RTN\textbackslash future       &0.298\\
RTN\textbackslash ptf=$\Delta$ &0.151\\
RTN\textbackslash ptf=1.0      &0.147\\
RTN\textbackslash ptf=0.0      &0.109\\
RTN\textbackslash resnet       &0.114\\
RTN\textbackslash h0           &0.095\\
RTN\textbackslash hcommon      &0.092\\
RTN                            &\textbf{0.087}\\
\end{tabular}
\end{center}
\end{sc}
\label{table:ablation}
\end{table}

The \begin{sc}hcommon\end{sc} and \begin{sc}h0\end{sc} modifications correspond to using a common, learned initial hidden state and null initial state for the LSTM layer respectively. The \begin{sc}resnet\end{sc} modification indicates that the RTN outputs the next positions instead of the offset from the previous ones. The \begin{sc}ptf\end{sc} modifications relate to the probability of the teacher forcing sampling, where the number corresponds to the probability of using the previous true frame, while the $\Delta$ symbol corresponds to the scheduled sampling strategy, where $p$ linearly decreases from $1$ to $0$ during training. The \begin{sc}future\end{sc} variant of the RTN corresponds to removing the future-context conditioning from the system. This is unsurprisingly the most necessary component for transition generation, and highlights the benefits of our LSTM modifications. Our proposed technique for the LSTM hidden initialization also improves generation and outperforms two commonly used techniques by allowing the hidden state to be initialized with respect to the treated sequence. In summary, Table 2 shows that our different design choices for the RTN all improve the general performance.   

\subsection{Temporal Super-Resolution}\label{section:superres}
In these final experiments, we explore one of the possible application of the RTN beyond transition generation. We use our system to decompress animation saved with only 10 frames of context and a single target state (pose + velocity) per second. The RTN is then responsible for doing temporal super-resolution of the animation. For the exploration of such an application, we train here an RTN on our whole dataset, without terrain conditioning and we test this approach on random validation sequences. In this setup, the network runs with the first 10 frames of animation as past context and the first target as future context. Afterwards, we can run the network in series, using its own generated transition as past context for the next transition animation. We apply the target blend postprocess to each generated second of transition, resulting in smooth and realistic motions and high-quality lossy decompression. We show such a decompressed sequence in Figure \ref{fig:superres}. We further suggest the reader to see the supplementary video to visualize longer decompressed sequences.
\begin{figure}[h]
\begin{center}
\centerline{\includegraphics[width=0.49\textwidth]{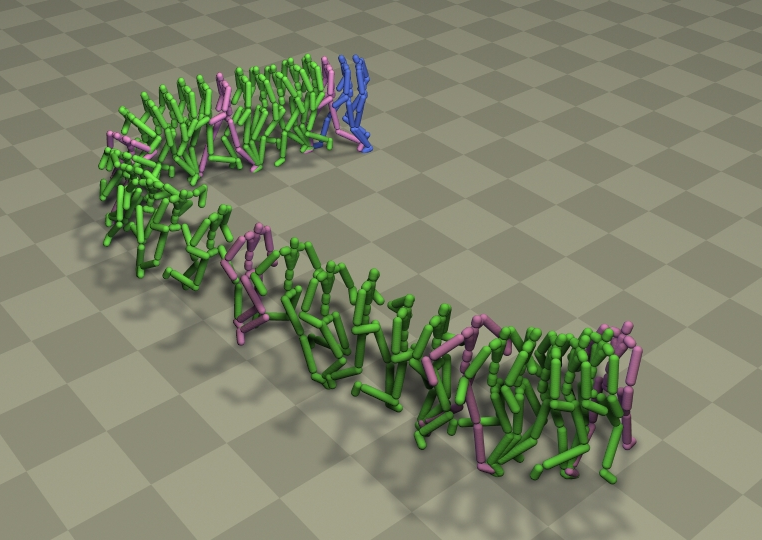}}
  \caption{An example of temporal super-resolution produced by the RTN. The saved animation only contains the initial context (blue) and a single target (magenta) per second. The network is able to generate realistic transitions, successfully performing temporal decompression, or super-resolution. Every fifth frame is shown. Full sequences are shown in the accompanying video.}
  \label{fig:superres}
\end{center}
\end{figure}
Some high frequency motions may be absent from the reconstruction, such as punches or dance moves, but this is due to our naive fixed-interval sampling of the keyframes. One could always chose which targets to store with one-second-or-less intervals in order to ensure the reach of important action landmarks.

\section{Limitations}
Our system, like other machine learning methods, is limited by the data it is trained on and requires large amount of quality samples and a computationally expensive learning phase. Another limitation of our system, like most of the related deep-learning based approaches for motion generation, is that it is not a probabilistic model and thus doesn't allow sampling of multiple transitions given the same constraints and can't model uncertainty well, as opposed to the Gaussian Process based methods. 

A limitation specific to our system emerges from the standardization of the global offset vector $\bo_t$. This standardization effectively limits the ability of the network to generate transitions of lengths significantly longer than those on which the statistics were computed. No architectural choice limits the length of the transitions to be generated, but the length on which the standardizing statistics are computed impose a soft upper bound on the duration of the producible transitions.

\section{Conclusion}
In conclusion, we propose a data-driven approach for animation transition generation, based on a custom recurrent architecture built on recent advances in deep learning for motion prediction and novel additions, such as a future-aware LSTM layer and an effective hidden state initialization procedure. Coupled with good input representations, such as future-context vectors and terrain local patches, we show robust results from a generic model for a wide range of motions. We hope that such a system will replace some naive interpolation strategies commonly used in animation authoring tools or lighten animation graphs by effectively replacing certain transitions nodes. We further explore applications by showing high-quality results on a temporal super-resolution task, where the network produces realistic motions from highly compressed animations.

Future work could include exploring bi-directional recurrent layers for replacing the target-blend postprocess, but such an approach would require nearly twice the computation time and memory requirements, as the recurrent layer is by far the biggest one, and would make other applications such as the super-resolution or other control-related tasks less appealing or impossible since it would require future contexts of several frames instead of single targets. Other directions could be to explore conditioning on other constraints than terrain, such as style or emotion for more refinement of the motions. Probabilistic deep-learning approaches such as variational auto-encoders or adversarial networks should also be explored for adding sampling capabilities to such networks.




%
%
%

\begin{acks}
We first thank Daniel Holden for his important inputs and suggestions. We also thank Simon Clavet, Mike Yurick, Julien Roy and David Kanaa for their help as well. Finally, we thank Ubisoft Montreal and the Mitacs Accelerate program (IT11395) for supporting this research.
\end{acks}




\break
\appendix
\section{Other dataset configurations}\label{app:A}
Tables \ref{table:samples} and \ref{table:windows} show differences in the data used for the experiments depending on the size of the dataset and the desired transition length.

\begin{table}[h]
\caption{Sample count for the datasets used.}
\begin{sc}
\begin{center}
\begin{tabular}{lrr}
Dataset &Training &Validation
\\ \hline
Terrain locomotion      &4 611 &1 008\\
All locomotion      &20 789 &4 468\\
\end{tabular}
\end{center}
\end{sc}
\label{table:samples}
\end{table}

\begin{table}[h]
\caption{Window specifications depending on the transition length.}
\begin{sc}
\begin{center}
\begin{tabular}{lrr}
Transition &Width &Offset
\\ \hline
1 second   &50 &20\\
2 seconds  &80 &20\\
\end{tabular}
\end{center}
\end{sc}
\label{table:windows}
\end{table}

\end{document}